\documentclass[aps,showpacs]{revtex4}
\usepackage{lscape,graphicx}
\usepackage{dcolumn}
\usepackage{bm}
\usepackage{epsfig}

\newcommand{\bea}{\begin{eqnarray}}
\newcommand{\eea}{\end{eqnarray}}
\newcommand{\beq}{\begin{equation}}
\newcommand{\eeq}{\end{equation}}

\begin{document}

\title{Low energy $0^+$ excitations in $^{158}$Gd}

\author{Jorge G. Hirsch}
 \affiliation{Instituto de Ciencias
Nucleares, Universidad Nacional Aut\'onoma de M\'exico, Apartado
Postal 70-543 M\'exico 04510 DF, M\'exico }

\author{Gabriela Popa, Shelly R. Lesher\footnote{currently at K.U. Leuven, Instituut voor Kern- en Stralingsfysica, Celestijnenlaan 200D, B-3001 Leuven, Belgium}  
and Ani Aprahamian}
\affiliation{Department of Physics, University of Notre Dame,
Notre Dame, IN 46556 , USA}

\author{Carlos E. Vargas} 
\affiliation{Facultad de F\'\i sica e Inteligencia Artificial, Universidad Veracruzana,
Sebasti\'an Camacho 5, Xalapa, Ver. 91000, M\'exico}

\author{Jerry P. Draayer}
 \affiliation{Department of Physics and Astronomy, Louisiana State University, 
Baton Rouge, Louisiana 70803, U.S.A}

\date{\today}
\begin{abstract}
High precision (p,t) studies of the deformed nucleus $^{158}$Gd allowed the 
observation of 13 excited $0^+$ states below an excitation energy of 3.1 MeV. This high 
density of low energy states, and particularly their measured B(E2) transition strengths 
to the first excited $2^+$ state challenge nuclear models.
The pseudo SU(3) model, which successfully describes many excited bands in Dy isotopes,
is used to analyze this nuclei. We have fairly good reproduction of most of the states but the absence
of actively including nucleons occupying intruders orbits may be the reason for the observed
limitations of the model.
\end{abstract}

\maketitle

\section{Introduction}

Recent experiments had provided clear evidence of the existence of many low lying 
$K^+ = 0^+$ bands in deformed nuclei. Using a Q3D spectrometer in high-precision (p,t) 
studies 13 excited $0^+$ states have been identified below 3.1 MeV in $^{158}$Gd, seven 
of them for the first time \cite{Lesh02}. The new $0^+$ assignments were strengthened by 
the placement of $\gamma$ rays that were identified to belong to the $^{158}$Gd 
nucleus with no previous level assignments. Such an abundance of $0^+$ states had not 
previously been seen in nuclei.

This high density of $0^+$ states challenges the simplest theoretical descriptions, which 
usually predict very few bands below 3 MeV. This is true for the earlier studies using the 
geometrical collective model (GCM) \cite{Tro91}, and the {\em sd-}IBM \cite{Iach87}, which 
could account for only 5 excited $0^+$ states below 3 MeV. The inclusion of the octupole 
degree of freedom in IBA calculations allows the prediction of 10 excited $0^+$ states 
below 3 MeV and 14 below 4 MeV, a number of them having a strong collective two-phonon 
octupole character \cite{Zam02}. The Projected Shell Model (PSM) \cite{Har95}, using as 
building blocks angular-momentum-projected two- and four- quasiparticle (qp) states, is 
also able to reproduce reasonably well the energies of all observed $0^+$ states 
\cite{Sun03}. Most of the states are dominated by one 2-qp or 4-qp state, coming from the 
near-Fermi Nilsson levels with low excitation energies. The qp nature of these excited 
states is quite different from the collective octupole vibration introduced in \cite{Zam02}. 

A microscopic calculation within the quasiparticle-phonon model (QPM) \cite{Sol92} offered 
a less biased criterion for determining the nature of these $0^+$ states. In this model 
the microscopic phonons, both collective and noncollective, are generated in the 
random-phase approximation (RPA), and are used to diagonalize a separable Hamiltonian 
containing different multipoles. The study of the $0^+$ states in $^{158}$Gd found many 
low energy states, which in most cases have large, if not dominant, two-phonon octupole 
components \cite{LoI04}, in agreement with the previous IBA predictions \cite{Zam02}.

Measured electromagnetic transition strengths are extremely useful to distinguish between 
different theoretical models.
A strong $B(E2:K=0^+_2\rightarrow \gamma )$ had been determined in $^{158}$Gd using the 
GRID technique \cite{Bor99}.The authors associated this with band mixing rather than with a 
double-$\gamma$ phonon. Other $B(E2)$ values for the decay of the $K=2^+_1$ and $K=0^+_2$ 
bands were also reported in \cite{Bor99}. The study of $0^+$ excitations with the 
$(n, n' \gamma)$ reaction allowed the determination of $B(E2:K=0^+_x\rightarrow 2_1 )$ 
values for ten of the previously measured excited $0^+$ states \cite{Lesh02b}. They all 
have values between one and a few W.u., suggesting the presence of a significant 
fragmentation of the collectivity in these excited states. 

\begin{table}
$$
\begin{array}{cccc}
E_\gamma (keV)  & E_x (keV)  &J^\pi_x &B(E2) (W.u.)  \\~~\\
1116.48   & 1196.10 &0^+_2 & 2.3\pm 0.9 \\
1372.90   & 1452.30 &0^+_3 & 3.1\pm 1.1 \\
1878.3   & 1957.8 &0^+_7 & 4.1\pm 1.9 \\
2196.61   & 2277.0 &0^+_8 & 4.2\pm 1.6 \\
2260.47   & 2338.0 &0^+_9 & 1.0\pm 0.3 \\
2564.67   & 2643.4 &0^+_{10} & 6.4\pm 3.7 \\
2605.8   & 2687.1 &0^+_{11} & 4.5\pm 3.5 \\
2832.0   & 2911.2 &0^+_{12} & 0.9\pm 0.9 \\
2997.0   & 3076.7 &0^+_{13} & 1.7\pm 0.5 \\
3026.2   & 3109.9 &0^+_{14} & 1.9\pm 0.6 \\
\end{array}
$$
\caption{$\gamma$-ray energies, excitation energies of the state $J^\pi_x$ and the extracted 
$B(E2:K=0⁺_x\rightarrow 2_1 )$ values in $^{158}$Gd, taken from Ref. \cite{Lesh02b}}.
\end{table}

The theoretical description of these large but fragmented B(E2) values has not been 
possible up to now. From the 18 $0^+$ states predicted in the PSM to have energies below 
3.25 MeV \cite{Zam02}, only two states have B(E2) transition strengths larger than 1 W.u. 
In the QPM calculation only the first excited $0^+$ state is predicted to decay with a 
$B(E2)$ strength larger than 1 W.u.. 

In the present contribution we report on an attempt to describe the observed $0^+$ excited 
states and their $B(E2)$ transition strengths in $^{158}$Gd using the pseudo SU(3) model 
\cite{Hech69,Ari69,Rat73}. We were strongly motivated by the success of this model in 
describing the energy levels and electromagnetic transition strengths in many excited 
bands in $^{157}$Gd, $^{163}$Dy and $^{169}$Tm \cite{Var02}. The same model allowed also 
the description of up to 8 rotational bands in  $^{158,160,162,164}$Dy \cite{Var04}. In 
what follows we will briefly review the main ideas behind the pseudo SU(3) model, present 
the results for $^{158}$Gd, and discuss them critically. 

\section{Pseudo SU(3) basis}

\begin{figure}
\epsfxsize=9.00cm
\centerline{\epsfbox{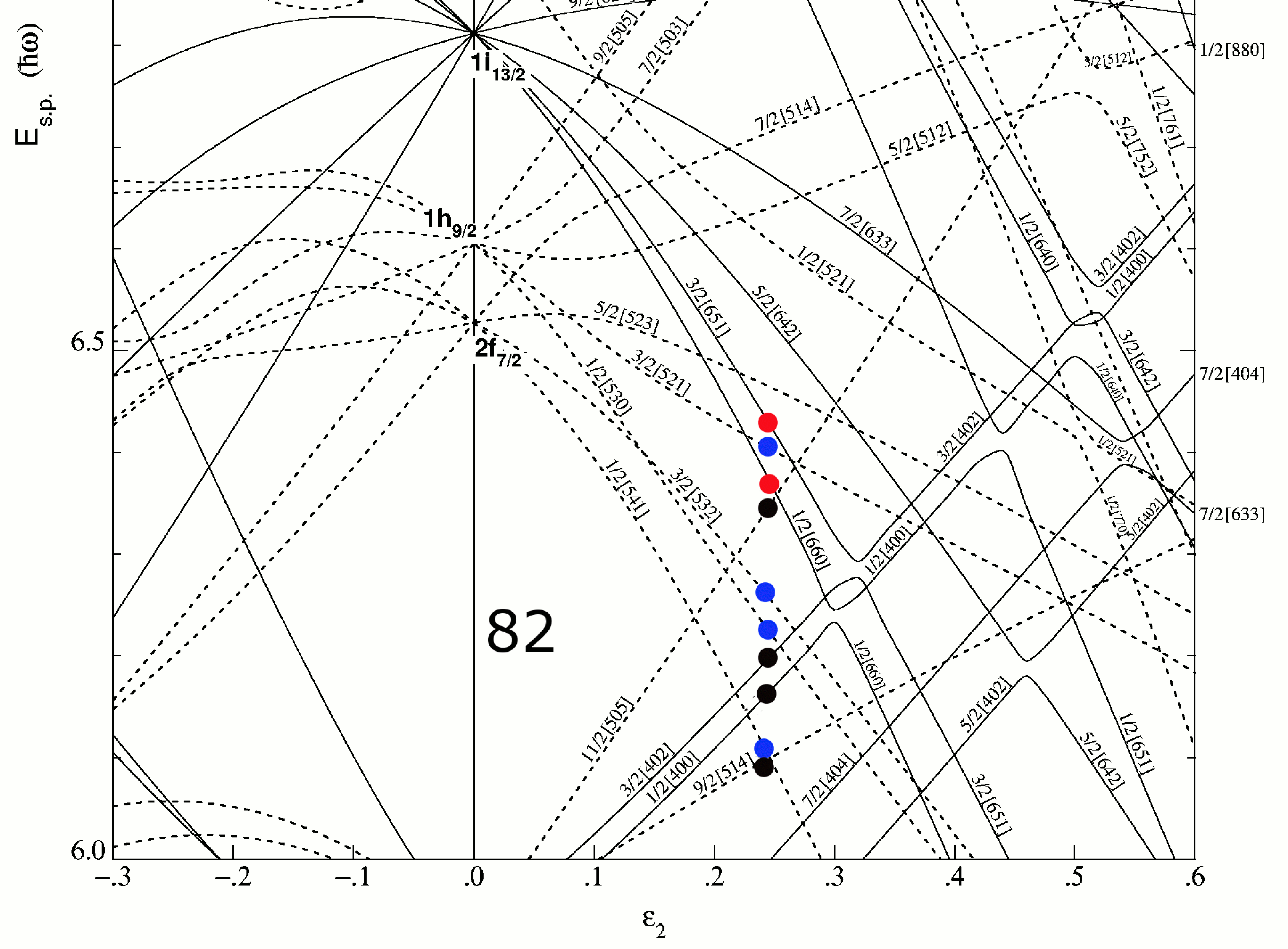}}
\centerline{\epsfbox{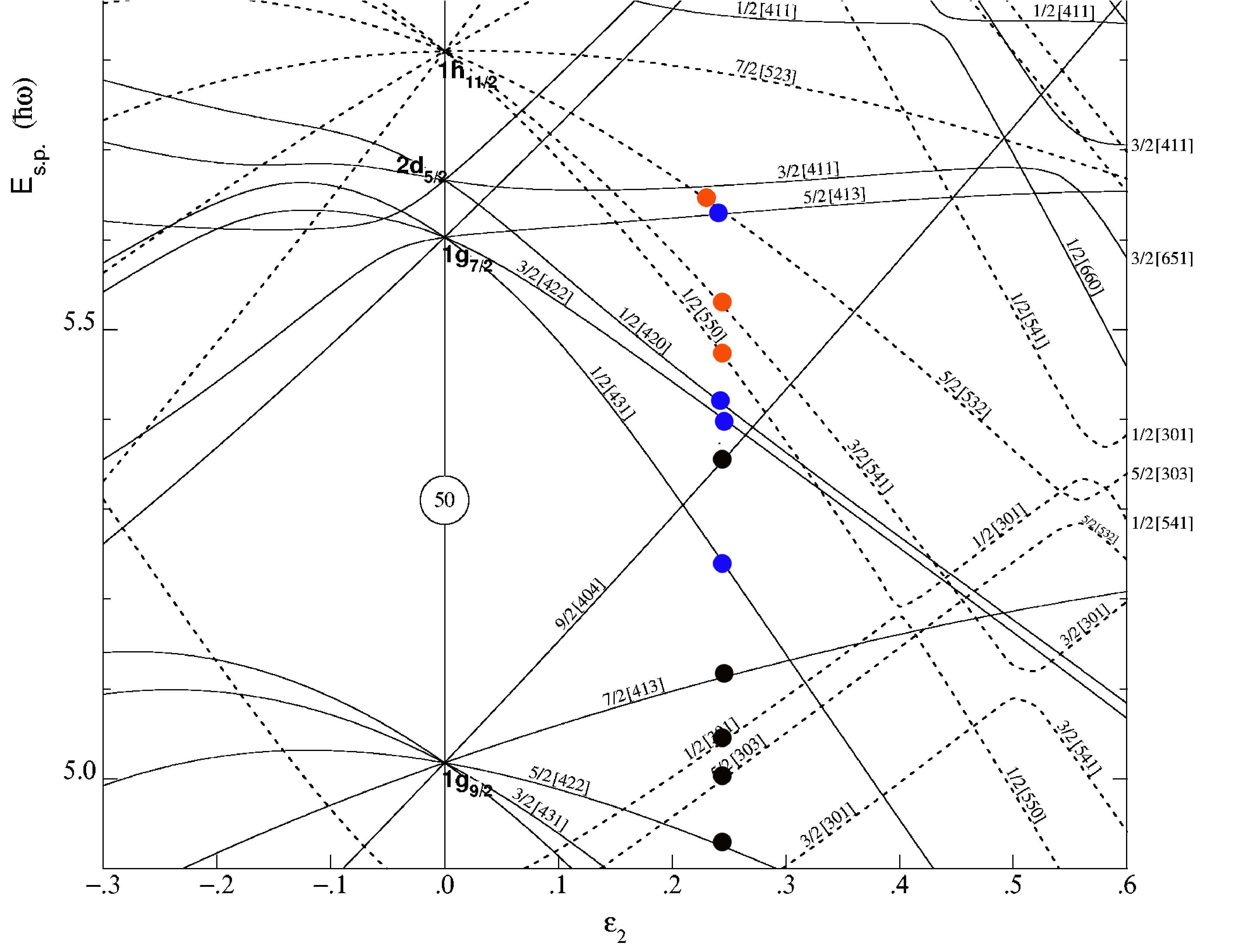}}
\caption{The occupancies for neutrons (top) and protons (bottow) as determined by
the filling of the the deformed Nilsson levels}
\label{occup}
\end{figure}

The pseudo SU(3) model  \cite{Dra82} has been widely used in recent years in the description of even-even 
\cite{Beu00,Pop00,Dra01} and odd-mass nuclei \cite{Var00a,Var00b,Var01}.
The first step in any application of the pseudo SU(3) model is to build the
many-body basis. For the pseudo SU(3) scheme the proton and neutron valence  
Nilsson single-particle levels are filled from below for a fixed deformation,
which in the case of $^{158}$Gd is $\epsilon_2 = 0.25$ \cite{Moll95}.
It allows the determination of the most probable normal and unique parity
orbital occupancies, as shown in Fig. \ref{occup}. 
Of the 14 valence protons, 8 occupy normal parity orbitals, and 6 intruder orbitals.
Of the 12 valence neutrons, 8 are in normal parity orbitals and  4 in intruder orbitals.

Many-particle states are built as pseudo-SU(3) coupled states with a well-defined particle 
number (of nucleons in normal parity orbits) and good total angular momentum. 
Nucleons occupying the intruder orbits are considered
implicitly through the use of effective charges. The explicit
inclusion of the unique-parity sector configurations remains an
open challenge that, while under investigation, is still not
available.  

Since in a quadrupole-quadrupole driven Hamiltonian, the states corresponding
to highest deformation are the most important, we extract from
this scheme the proton and neutron  SU(3) irreps corresponding to
the highest $C_2$ values which, in turn, are coupled to final
SU(3) irreps that have good total angular momentum
\cite{Var00a,Dra82}. The configuration space was
generated from the strong coupling of the eight protons and eight
neutrons in the normal parity states. Two sets of calculations
were performed. In the first case proton and neutron states with
only pseudo-spin zero were considered and, in the second case, the
configuration space was enlarged by considering proton and
neutron states with both pseudo-spin zero and one. In both cases
the configuration space was truncated by considering all the
coupled SU(3) irreps with a $C_2$ greater than the same value for
the $C_{2 cut}$. The SU(3) irreps considered in the two cases
are given in Table \ref{gd158_irreps}, and \ref{gd158g_irreps}.

\begin{table}[h]
\begin{tabular}{|cc|l|}
\hline $(\lambda_\pi,\mu_\pi)$ & $(\lambda_\nu,\mu_\nu)$ &
   $(\lambda,\mu)$ \\
\hline
(10,4)&(18,4)&(28,8), (29,6), (30,4), (31,2), (32,0), (26,9), (27,7)\\
(10,4)&(20,0)&(30,4)\\
(10,4)&(16,5)&(26,9),(27,7)\\
(10,4)&(17,3)&(27,7)\\
(10,4)&(13,8)&(23,12)\\
(12,0)&(18,4)&(30,4)\\
(12,0)&(20,0)&(32,0)\\
(8,5)&(18,4)&(26,9),(27,7)\\
(9,3)&(18,4)&(27,7)\\
(5,8)&(18,4)&(23,12)\\
 \hline
\end{tabular}
\caption{Case 1: The SU(3) irreps (obtained by coupling all the
pseudo-spin zero proton and neutron irreps) with $C_2 > C_{2
cut}$,  ordered by decreasing $C_2$ values, used to describe
the low-energy spectra in $^{158}$Gd.} \label{gd158_irreps}
\end{table}

\begin{table}[h]
\begin{tabular}{|cc|cc|cc|c|cc|cc|cc|}
\hline $(\lambda_\pi,\mu_\pi)$ &$S_\pi$&  $(\lambda_\nu,\mu_\nu)$
& $S_\nu$ & $(\lambda,\mu)$ & $S$ &~~~
& $(\lambda_\pi,\mu_\pi)$ &$S_\pi$&  $(\lambda_\nu,\mu_\nu)$
& $S_\nu$ & $(\lambda,\mu)$ & $S$   \\
\hline
(10,4)&0&(18,4)&0& (28,8)&0 &&
(10,4)&0&(18,4)&0& (29,6)&0\\
(10,4)&0&(19,2)&1& (29,6)&1 &&
(11,2)&1&(18,4)&0& (29,6)&1\\
(10,4)&0&(18,4)&0& (30,4)&0 &&
(10,4)&0&(19,2)&1& (30,4)&1\\
(10,4)&0&(20,0)&0& (30,4)&0 &&
(11,2)&1&(18,4)&0& (30,4)&1\\
(11,2)&1&(19,2)&1& (30,4)&0 &&
(11,2)&1&(19,2)&1& (30,4)&1\\
(11,2)&1&(19,2)&1& (30,4)&2&&
(12,0)&0&(18,4)&0& (30,4)&0\\
(10,4)&0&(18,4)&0& (31,2)&0&&
(10,4)&0&(19,2)&1& (31,2)&1\\
(11,2)&1&(18,4)&0& (31,2)&1 &&
(11,2)&1&(19,2)&1& (31,2)&0\\
(11,2)&1&(19,2)&1& (31,2)&1 &&
(11,2)&1&(19,2)&1& (31,2)&2\\
(11,2)&1&(20,0)&0& (31,2)&1 &&
(12,0)&1&(19,2)&1& (31,2)&1\\
(10,4)&0&(18,4)&0& (32,0)&0 &&
(11,2)&1&(19,2)&1& (32,0)&0\\
(11,2)&1&(19,2)&1& (32,0)&1 &&
(11,2)&1&(19,2)&1& (32,0)&2\\
(12,0)&0&(20,0)&0& (32,0)&0 &&
(10,4)&0&(18,4)&0& (26,9)&0\\
(10,4)&0&(16,5)&0& (26,9)&0 &&
(10,4)&0&(16,5)&1& (26,9)&1\\
( 8,5)&0&(18,4)&0& (26,9)&0 &&
( 8,5)&1&(18,4)&0& (26,9)&1\\
(10,4)&0&(18,4)&0& (27,7)&0 &&
(10,4)&0&(19,2)&1& (27,7)&1\\
(10,4)&0&(16,5)&0& (27,7)&0 &&
(10,4)&0&(16,5)&1& (27,7)&1\\
(10,4)&0&(17,3)&0& (27,7)&0 &&
(11,4)&0&(17,3)&1& (27,7)&1\\
(11,2)&1&(18,4)&0& (27,7)&1 &&
(11,2)&1&(16,5)&0& (27,7)&1\\
(11,2)&1&(16,5)&1& (27,7)&0 &&
(11,2)&1&(16,5)&1& (27,7)&1\\
(11,2)&1&(16,5)&1& (27,7)&2 &&
( 8,5)&0&(18,4)&0& (27,7)&0\\
( 8,5)&1&(18,4)&0& (27,7)&1 &&
( 8,5)&0&(19,2)&1& (27,7)&1\\
( 8,5)&1&(19,2)&1& (27,7)&0 &&
( 8,5)&1&(19,2)&1& (27,7)&1\\
( 8,5)&1&(19,2)&1& (27,7)&2 &&
( 9,3)&0&(18,4)&0& (27,7)&0\\
( 9,3)&1&(18,4)&0& (27,7)&1 &&
(10,4)&0&(13,8)&0&(23,12)&0\\
(10,4)&0&(13,8)&1&(23,12)&1 &&
( 5,8)&0&(18,4)&0&(23,12)&0\\
( 5,8)&1&(18,4)&0&(23,12)&1 &&&&&&& \\
\hline
\end{tabular}
\caption{Case 2: The SU(3) irreps (obtained by coupling all the
pseudo-spin zero and one proton and neutron irreps) with $C_2 >
C_{2 cut}$, the same as in the first case, ordered by
decreasing $C_2$ values, used to describe the low-energy spectra in
$^{158}$Gd.} \label{gd158g_irreps}
\end{table}

Any state $| J_i M \rangle$, where $J$ is the total angular momentum, $M$ its
projection and $i$ an integer index which enumerates the states with the same
$J, M$ starting from the one with the lowest energy, is built as a linear
combination
\begin{equation} 
| J_i M \rangle = \sum_\beta C^{Ji}_\beta |\beta JM \rangle \label{wf}
\end{equation}  
of the strong coupled proton-neutron states
\begin{widetext}
\begin{eqnarray}
|\beta JM \rangle & \equiv &
| \{ \tilde{f}_\pi \} (\lambda_\pi \mu_\pi) \tilde{S}_\pi, \{ \tilde{f}_\nu \}  
(\lambda_\nu \mu_\nu) \tilde{S}_\nu ; \rho (\lambda \mu ) \kappa \tilde{L},\tilde{S}~ 
JM \rangle~~~~~~~~~~~~~~~~~~~~~~~~~~~~ \nonumber \\
 & = & \sum_{M_L M_S} (\tilde{L} M_L, \tilde{S} M_S | J M ) \sum_{M_{S \pi} M_{S \nu}}
(\tilde{S}_\pi M_{S \pi}, \tilde{S}_\nu M_{S \nu} | \tilde{S} M_S)  \nonumber \\
& & \sum_{k_\pi \kappa_\nu \tilde{L}_\pi \tilde{L}_\nu M_\pi M_\nu }
{ \langle (\lambda_\pi \mu_\pi) \kappa_\pi \tilde{L}_\pi M_\pi ;
(\lambda_\nu \mu_\nu) \kappa_\nu \tilde{L}_\nu M_\nu |
(\lambda \mu ) \kappa \tilde{L} M \rangle}_\rho \label{basis} \\
& & 
| \{ \tilde{f}_\pi \} (\lambda_\pi \mu_\pi) \kappa_\pi \tilde{L}_\pi M_\pi, 
\tilde{S}_\pi M_{S \pi}  \rangle
| \{ \tilde{f}_\nu \} (\lambda_\nu \mu_\nu) \kappa_\nu \tilde{L}_\nu M_\nu, 
\tilde{S}_\nu M_{S \nu} \rangle  .\nonumber
\end{eqnarray}
\end{widetext}
In the above expression $(-,-|-)$ and $\langle-;-|- \rangle$ are the SU(2) and
SU(3) Clebsch Gordan coefficients, respectively.  

In this article we consider the Hilbert space spanned by the states with $\tilde{S}_{\pi,\nu}$ = 0 and 1 in Eq. (\ref{basis}). The main difference 
with the pseudo SU(3) basis used in previous pseudo SU(3) descriptions of even-even nuclei \cite{Pop00} is the inclusion of states with  $\tilde{S}_{\pi,\nu}$ = 1 in the proton and neutron wave functions. They have a non negligible contribution to excited rotational bands. 
The goodness of the pseudo SU(3) symmetry is preserved by imposing that states with  $\tilde{S}_{\pi,\nu}$ = 0 should be dominant in the ground state. It translates into severe limits for the ``rotor-like" terms in the Hamiltonian, and guarantees that the whole band structure is preserved.

\section{The Hamiltonian}

The Hamiltonian has a {\it principal} part $H_0$:
\begin{eqnarray}
H_0 & = & \sum_{\alpha=\pi,\nu} \{ H_{sp,\alpha} - G_\alpha 
~H_{pair,\alpha} \} - \frac{1}{2}~  \chi~ \tilde Q \cdot \tilde Q ~.
\label{eq:h0}
\end{eqnarray}
which contains spherical Nilsson single-particle terms for protons 
and neutrons ($H_{sp,\pi[\nu]}$), quadrupole-quadrupole ($\tilde Q \cdot \tilde 
Q$) and pairing ($H_{pair,\pi[\nu]}$) interactions. 
Added to these are five `rotor-like' terms that are diagonal in the SU(3) 
basis:
\begin{eqnarray}
    H  =  H_0 + a J^2 + b K_J^2 + a_3 C_3 + a _{sym} C_2 + d S^2. \label{eq:ham}
\end{eqnarray}
A detailed analysis of each term of this Hamiltonian and its parameterization can be found 
in \cite{Var00a}. The different terms in $H_0$  have been widely studied in the nuclear 
physics literature, allowing their respective strengths to be fixed by systematics 
\cite{Var00a,Rin79,Duf96}. The configuration mixing is due to the SU(3) symmetry-breaking 
Nilsson single-particle and pairing terms. 

The single-particle terms ($H_{sp,\alpha}$) have the form:
\begin{eqnarray}
H_{sp,\alpha} = \sum_{i_\alpha} \left(C_\alpha {\bf  l}_{i_\alpha} \cdot 
{\bf s}_{i_\alpha} + D_\alpha {\bf  l}^2_{i_\alpha}\right),    
~~~~~\alpha=\pi,\nu,
\end{eqnarray}
where $C_\alpha$ and $D_\alpha$ are fixed following the usual prescriptions \cite{Rin79}. 
In the pseudospin basis the spin-orbit 
and orbit-orbit contributions are small, but they still generate most of the mixing between 
pseudo SU(3) irreps. 

The `rotor-like' terms in Hamiltonian (\ref{eq:ham}) are used to fine tune the calculated 
spectra. The five parameters $a, b, a_3, a_{sym}, d$ were fixed following the prescriptions 
given in Ref. \cite{Pop00,Var00a}. The $K_J^2$ breaks the SU(3) degeneracy of the different 
K bands, the $J^2$ term provides small corrections to the moment of inertia. These two 
terms help to fit the energy of the $\gamma$ band and the moment of inertia of the ground 
band, respectively. 
It is worth keeping in mind that these two terms only modify the wave function slightly,  
their main effects is on the energies.

The parameters of $a_{sym}$ and $a_3$ in the $\tilde C_2$ and $\tilde C_3$ terms
must be strongly restricted to avoid drastic changes in the wave functions.
The theory is most sensitive to the parameter $\tilde C_3$, because when large values for 
$c$ are employed, the ground state becomes a pure pseudospin 1 state.
It can also induce an artificially triaxial ground state in a well deformed nuclei.

\section{Results}

With all the fitting parameters set to zero in the Hamiltonian, 
we have poor agreement with the energies of the observed 0+ states,
but we do get all 13 0+ states. Also, the 1+ state is calculated at 620 keV
when it is observed at 1.84 MeV. In this case the first two excited $0^+$ 
states are lower in energy than their experimental counterparts. 
The first excited $0^+$ state changes very little with $a_{sym}$ parameter 
for a fixed value of the $a_3$. By increasing the $a_3$ 
parameter we fit the energy of the first excited $0_+$ state.
By increasing the $a_{sym}$ parameter, many states are being pushed higher in 
energy, including the excited $0^+$ states.  
By varying the coefficient in front of $K_J^2$, the energy value of the $K=2^+$ state
can be fit. 

\begin{figure}[th]
\centerline{\hbox{\epsfig{figure = 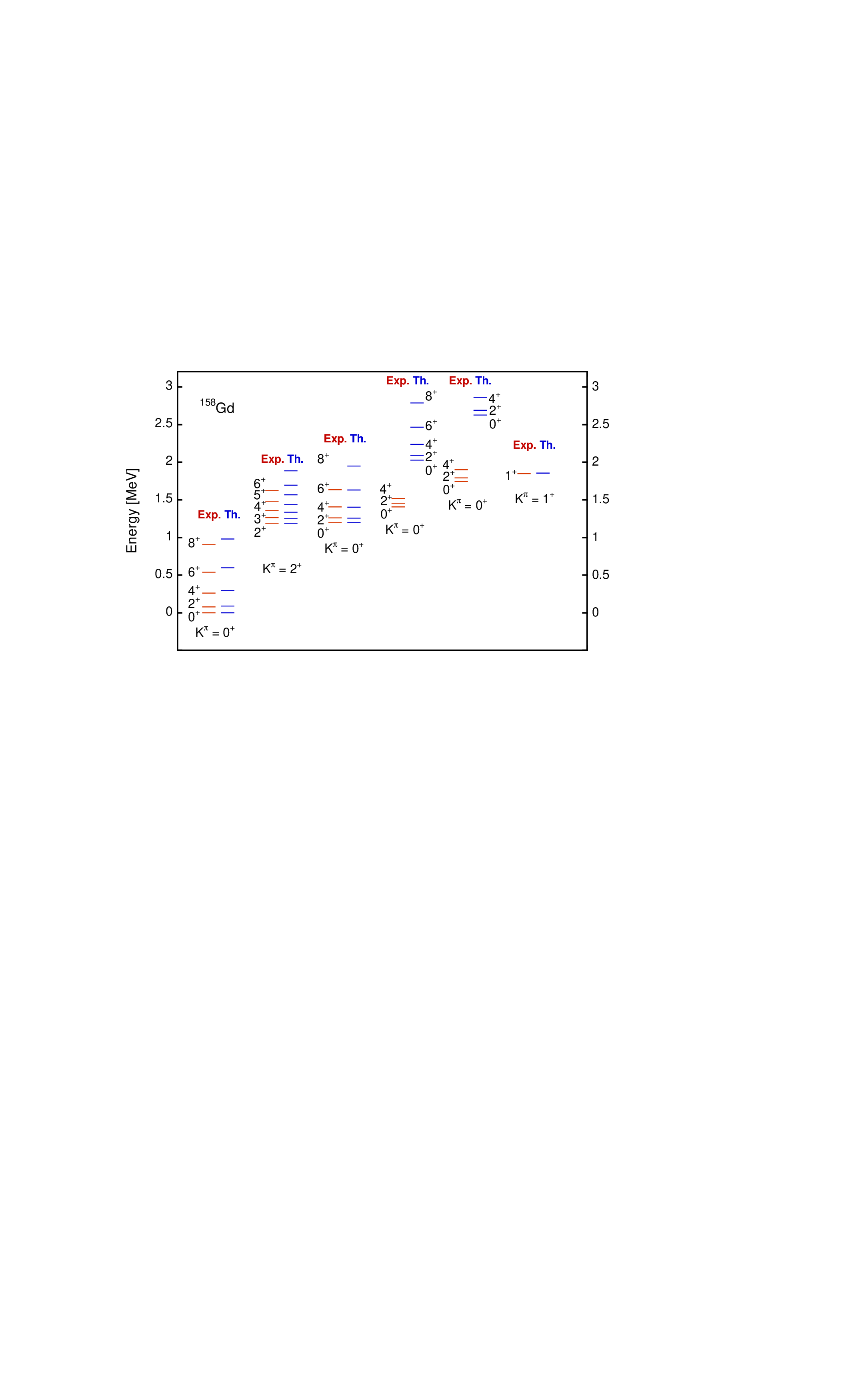,
width=16cm}}}
\caption{Case 1: Low energy spectra of $^{158}$Gd
obtained with the restricted configuration space. The experimental
values are shown on the left-hand side of each band (red lines), and the calculated ones 
on the right-hand side (blue lines). } \label{gd158en}
\end{figure}

With the set of parameters we determined, we were able to identify the low 
energy spectra and compare it with the experimental one in Fig. \ref{gd158en}.
As can be seen, the rotor-like terms in the Hamiltonian allowed for the adjustment of 
the moment of inertia of the ground-state band, and the energies of the $2^+_\gamma$, 
$0_2$ and $1^+_1$ states. It is clear, however, that the predicted third and fourth $0^+$ 
states have higher energies than their experimental counterparts, and that there are only 
4 $0^+$ states below 3 MeV.

\begin{figure}[th]
\centerline{\hbox{\epsfig{figure = 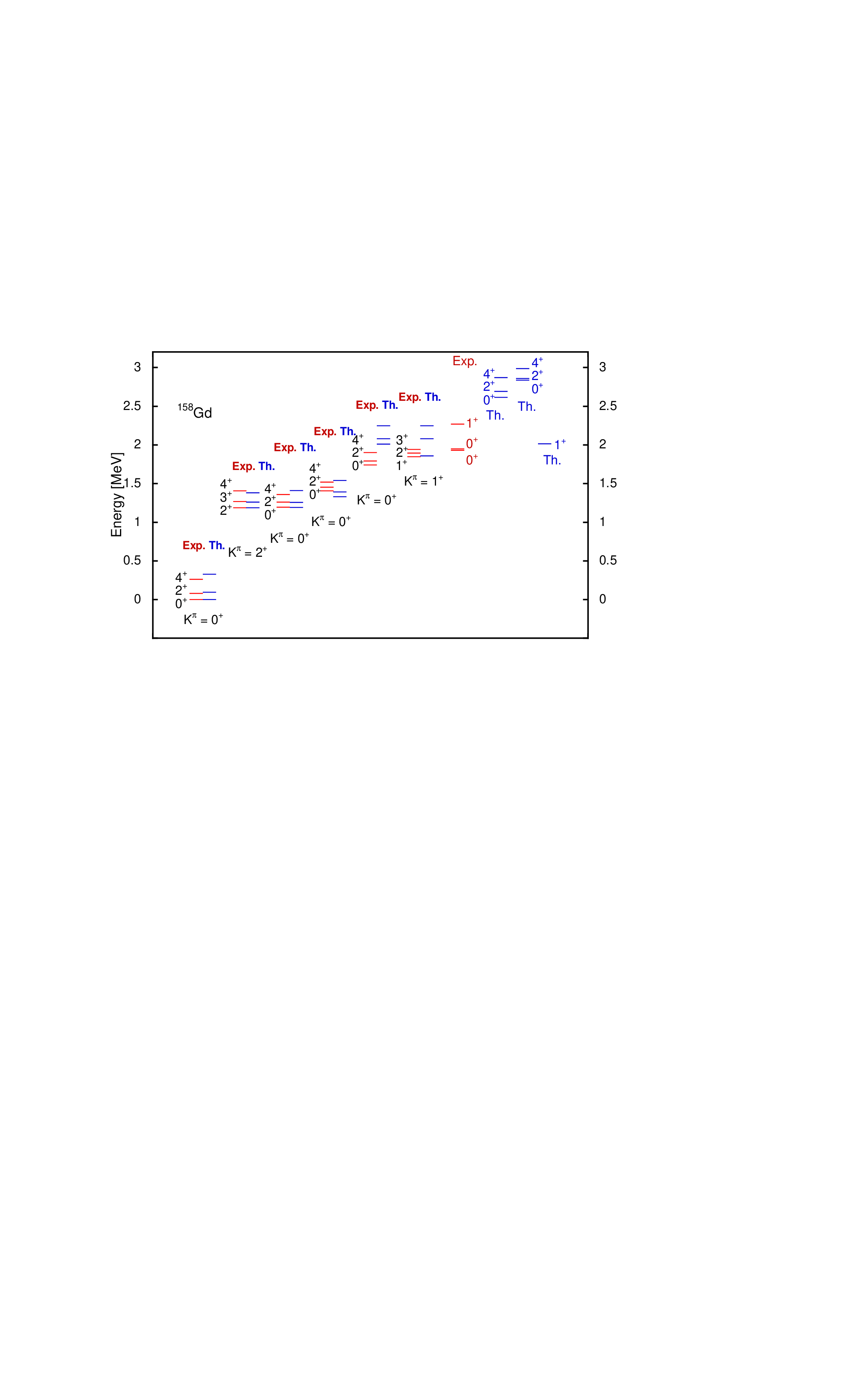,
width=16cm}}}
\caption{Case 2: Low energy spectra of $^{158}$Gd
obtained with an enlarged configuration space. Conventions as in Fig. \ref{gd158en}. } 
\label{gd158enb}
\end{figure}

Using the enlarged basis listed in Table \ref{gd158g_irreps} we obtained the low energy 
spectra shown  in Fig. \ref{gd158enb}, which is also compared with the experimental levels.
The positive effects of enlarging the basis are clearly seen in this figure. With the same 
number of parameters, there are now 7 $0^+$ states below 3 MeV, whose energies are mostly 
close to the measured ones.
This results is not suprising, because both in the IBM and in the PSM the use of enlarged 
basis allowed for the description of many $0^+$ states at low excitation energy.
 
A detailed analysis of the excited $0^+$ states wave functions, and their inter-band B(E2) 
transition strengths will be reported elsewhere \cite{Pop05}. In the present contribution 
we restrict the discussion to the B(E2) transition strengths between the excited  $0^+$ 
states and the $2_1^+$ state belonging to the ground state band, and the $2_\gamma^+$ state, 
the $\gamma$-bandhead. These B(E2) transition strengths are listed in Table \ref{be2} for 
the small basis, case 1, and in Table \ref{be2l} for the large basis, case 2. In both tables
the excitation energy of the $0_i$ state is given in the first column, the calculated 
$B(E2; 0_i \rightarrow 2_1)$ in the second column, and the calculated $B(E2; 0_i 
\rightarrow 2_\gamma)$ in the third column.

\begin{table}[h]
\begin{tabular}{|c|cc|}
\hline
Energy & $B(E2; 0_i \rightarrow 2_1)$ & $B(E2; 0_i \rightarrow 2_\gamma)$ \\
(MeV) & (W.u.) & (W.u.)\\
\hline
 1.19 & 0.08  & 1.22\\
 2.03 & 0.0004& 7.97\\
 2.62 & 0.001 & 2.58\\
 3.59 & 0.01  & 0.27\\
\hline
\end{tabular}\caption{Case 1: The energy value of each
excited $0^+$ states is given in the first column. The
corresponding transition values $B(E2;0_i-> 2_1) $  are given in
the second column, and the $B(E2; 0_i -> 2_\gamma)$ in the third
column. }\label{be2}
\end{table}

\begin{table}[h]
\begin{tabular}{|c|cc|}
\hline
Energy & $B(E2; 0_i \rightarrow 2_1)$ & $B(E2; 0_i \rightarrow 2_\gamma)$ \\
of $0_i [MeV]$ & [W.u.] & [W.u.]\\
\hline
 1.19 & 0.06 & 7.79\\
 1.33 &0.0002& $< 1.0 E^{-6}$\\
 2.01 &0.001& 9.74\\
 2.61 &0.0002& 3.02\\
 2.83 &$< 1.0 E^{-6}$& $< 1.0 E^{-6}$\\
 3.56 &0.01& 0.26\\
 3.68 &$< 1.0 E^{-6}$&$< 1.0 E^{-6}$\\
 3.77 &$< 1.0 E^{-6}$&$< 1.0 E^{-6}$\\
\hline
\end{tabular}\caption{Case 2: The energy value of each
excited $0^+$ states is given in the first column. The
corresponding transition values $B(E2;0_i-> 2_1) $  are given in
the second column, and the $B(E2; 0_i -> 2_\gamma)$ in the third
column. }\label{be2l}
\end{table}

In both cases the transition strengths to the ground state band are two or more orders of 
magnitude smaller than the experimental ones. There are, however, three or four  B(E2) values 
to the $\gamma$- bandhead larger than  1 W.u. This suggests that, while many $0^+$ states 
are described in the pseudo SU(3) model at the right excitation energy, their wave functions 
are missing some important elements. The mixing of different occupancies in the normal 
parity sector, induced by the pairing interactions \cite{Hir03}, could correct for these 
deficiencies.

\section{Conclusions}

The excitation energies of many $0^+$ states in $^{158}$Gd can be properly described using 
the pseudo SU(3) model including states with pseudo-spin 0, 1 and 2.
While the calculated B(E2) transition strengths to the g.s. band are smaller than the 
observed ones, those to the gamma band are of the same order, measured in W.u..
Calculations in $^{156}$Gd suggest that configuration mixing (different normal and intruder 
occupations mixed by pairing) could allow stronger transitions to the g.s. band.

\section*{Acknowledgments}
This work was supported in part by Conacyt (M\'exico), DGAPA-UNAM, and the U.S.
National Science Foundation, including the contract PHY01-40324.

\end{document}